# Universal superlattice potential for 2D materials from twisted interface inside h-BN substrate

Pei Zhao[1,2][#], Chengxin Xiao[1,2][#], Wang Yao[1,2][*]

[1] Department of Physics, University of Hong Kong, Hong Kong, China
[2] HKU-UCAS Joint Institute of Theoretical and Computational Physics at Hong Kong, China

[*] Correspondence to: wangyao@hku.hk

[#] These authors contributed equally.

**Abstract:**

Lateral superlattices in 2D materials are emerging as a powerful platform for exploring novel quantum phenomena, which can be realized through the proximity coupling in forming moiré pattern with another layer. This approach, however, is invasive, material-specific, and requires small lattice mismatch and suitable band alignment, largely limited to graphene and transition metal dichalcogenides (TMDs). Hexagonal boron nitride (h-BN) of anti-parallel (AA') stacking order has been an indispensable building block, as dielectric substrates and capping layers for realizing high quality van der Waals devices. There is also emerging interest on parallelly aligned h-BN of Bernal (AB) stacking, where the broken inversion and mirror symmetries lead to out-of-plane electrical polarization with sign controlled by interlayer translation. Here we show that the laterally patterned electrical polarization at a nearly parallel interface within the h-BN substrate or capping layer can be exploited to create non-invasively a universal superlattice potential in general 2D materials. The feasibility is demonstrated by first principle calculations for monolayer $MoSe_2$, black phosphorus, and antiferromagnetic $MnPSe_3$ on such h-BN substrate. The potential strength can reach 200 meV, customizable in this range through choice of vertical distance of target material from the interface in h-BN. We also find sizable out-of-plane electric field at the h-BN surface, which can realize superlattice potential for interlayer excitons in TMD bilayers as well as dipolar molecules. The idea is further generalized to AB stacked h-BN film subject to torsion with adjacent layers all twisted with an angle, which allows the potential and field strength to be scaled up with film thickness, saturating to a quasi-periodic one with chiral structure.

## Introduction

Lateral superlattices in graphene and transition metal dichalcogenides (TMDs) are drawing remarkable interest as new arena to explore emergent quantum transport,[1-4] quantum optical,[5-8] and many-body phenomena.[9-23] Such superlattices have been realized through forming long-wave moiré pattern, where the registry dependent proximity coupling from adjacent layers periodically modulate the electronic structures. This approach, however, requires the building blocks to have small lattice mismatch and suitable band alignment, and its implementation has so far been limited to graphene and TMDs. The resultant moiré superlattices are material-specific, and also dependent on the carrier species and internal degrees of freedom. Taking TMDs heterobilayers as example, the electrons at K and Q valleys and holes at K and Γ points all experience distinct moiré superlattice potentials, and the energy landscape depend on choices of compounds.[24,25] The proximity coupling at the moiré interface also degrade or change other material properties of the target layer. For example, the TMD heterointerface for realizing superlattice potential of electrons[18-23] also results in ultrafast electron-hole separation that completely alters the optical properties.[5-8]

The realization of the elegant moiré superlattices and high-quality van der Waals devices has relied on the use of hexagonal boron nitride (h-BN) as substrate and capping layer,[26,27] for its remarkable thermal and chemical stability, and insulating band structures. Commonly used h-BN has the anti-parallel alignment between adjacent layers, known as the AA' stacking, where boron atoms are vertically aligned with nitrogen atoms of neighboring layers. Another energetically favorable h-BN structure is to have the layers parallelly aligned with the AB (Bernal) stacking, and growth of such AB h-BN has been achieved using chemical vapor deposition.[28] Unlike the AA' one, the AB stacking allows the emergence of perpendicular electric polarization since the restriction from inversion symmetry and out-of-plane mirror symmetry are both lifted.[29] Observations of ferroelectricity in AB-stacked h-BN are reported very recently: under marginally twisting where structure relaxation leads to micron or sub-micron scale alternating AB and BA stacking domains, the opposite out-of-plane electrical polarizations are sensed by Kelvin probe force microscope,[30,31] and by the resistance of adjacent graphene.[32]

Here we propose a non-invasive approach to realize universal superlattice potential on a general 2D material, exploiting the electrical polarization at a nearly parallel interface inside the h-BN substrate or capping layer. Such interface can be readily created by stacking a *n*-layer AA' h-BN on another AA' film (Fig. 1), which hosts electrical polarization patterned with periodicity $b = a/\delta$, $a$ the lattice constant of h-BN and $\delta$ the twist angle from parallel. A triangular superlattice potential with period $b$ is then electrostatically generated in the target 2D material placed on h-BN surface, regardless of its crystal structure and orientation. This is corroborated by our first principle calculations for monolayers of MoSe$_2$, black phosphorus, and antiferromagnetic MnPSe$_3$ on h-BN substrates. The potential strength is proportional to $\exp\left(-nd\frac{4\pi}{\sqrt{3}b}\right)$, $nd$ the vertical distance of target material from the interface in h-BN, and can be customized in the range 0 - 200 meV. This rapid exponential dependence corresponds to sizable out-of-plane electric field $\sim 0.6 V/b$, which can realize superlattice potential for dipolar objects at the h-BN surface, such as interlayer excitons in TMDs bilayers, or adsorbed dipolar molecules. Using AB h-BN film subject to torsion where adjacent layers are all twisted with a common angle $\delta$, the strengths of potential and electric field on the surface can be further scaled up with film thickness $N$, which saturates to a quasi-periodic form with a chiral structure when $N\delta$ becomes sufficiently large.

**Results**

We first examine the electrical polarization at a nearly parallel interface in h-BN, with twist angle $\delta \ll \pi$ but beyond the marginally twisting regime addressed in Ref. [30-32] such that structural relaxation is not significant. This allows us to focus on a moiré pattern between rigid lattices at the nearly parallel interface. The local stacking registry can be characterized by the in-plane displacement $\boldsymbol{r}$ between two near-neighbor B atoms across the interface (Fig. 2a), which is a linear function of the lateral position $\boldsymbol{R}$. Having a few-layer h-BN to separate this interface from the surface may also help to stabilize the moiré from structural relaxation.

The nearly parallel interface is now described by a long-wavelength moiré pattern $\boldsymbol{r}(\boldsymbol{R})$, with period $b \gg a$. There exists an intermediate length scale ($l$), large compared to lattice constant $a$ and small compared to $b$, such that the atomic registry in any region of size $l$ closely resembles a lattice-matched stacking, so the local electronic structures in moiré can be well approximated by band structures of the latter.[33,34] First-principles calculations of various lattice-matched stackings can give the electrical polarization $P$ as function of registry $r$. Combined with the mapping between $r$ and location $R$ in the long-wavelength moiré, the laterally modulated electrical polarization can then be given as $P(R) \equiv P(r(R))$.

We start with first principle calculations of $n + n$ multilayers where the top $n$-layer AA' hBN is stacked on the bottom $n$-layer AA' hBN, with a lattice-matched parallel interface of various registry $r$. Fig. 2b shows an example of the charge redistribution caused by the interlayer coupling at the parallel interface with $r = a/\sqrt{3}$ (i.e. an AB interface). The presented quantity is the differential charge $\Delta\rho \equiv \rho_{n+n} - \rho_{n,t} - \rho_{n,b}$, where $\rho_{n+n}$ is the DFT calculated plane-averaged charge density of the $n + n$ multilayer, and $\rho_{n,t/b}$ is that of the pristine top (bottom) $n$-layer. Remarkably, for various thickness n considered, $\Delta\rho$ is predominantly distributed with nearly the same profile on the two layers constituting the parallel interface. This shows the charge redistribution is occurring at the parallel interface only, and the outer-layers in anti-parallel alignment only has very small effect on the resultant electrical polarization.

In the lattice-matched stacking, the spatially uniform electrical polarization corresponds to a voltage drop: $\Delta\varphi = \frac{1}{\varepsilon_0} P$, which can be directly extracted from the DFT calculations as the difference between the vacuum levels on the two side of the multilayer. The electrical polarization determined from this vacuum level difference, i.e. $\varepsilon_0 \Delta\varphi$, is shown in Fig. 2c (triangle symbols). The magnitude of $P$ is ~ 2 pC/m, consistent with other calculations for AB bilayers,[29] while the modest dependence on the layer thickness n can be attributed to the small screening effect by the outer-layers. Fig. 2c also shows the electrical polarization evaluated from the differential charge, $P = \int z\Delta\rho(z)dz$ (star symbols), which is consistent with the $\varepsilon_0 \Delta\varphi$ values.

Fig. 2d plots the electrical polarization as function of $r$ in a 2 + 2 configuration. $r = 2a/\sqrt{3}$ corresponds to an BA stacking interface, where $P$ is exactly the opposite of the AB stacking one. For the AA stacking ($r = 0$), the out-of-plane mirror symmetry forbids any polarization. In general, $r$ stacking registry is related to the $-r$ one by a spatial inversion, so $P$ is an odd function of $r$. Like many other quantities involving the interlayer processes,[25,35] the registry dependence can be well fitted by the lowest Fourier harmonics observing the rotational and translational symmetry,

$$P(r) = \frac{P_0}{9}\big(f_+(r) - f_-(r)\big), \quad f_\pm(r) \equiv \left| e^{i\mathbf{K}\cdot\mathbf{r}} + e^{i(\hat{C}_3 \mathbf{K}\cdot\mathbf{r} \pm \frac{2\pi}{3})} + e^{i(\hat{C}_3^2 \mathbf{K}\cdot\mathbf{r} \pm \frac{4\pi}{3})} \right|^2, \quad (1)$$

where $\mathbf{K}$ is wavevector at the Brillouin zone corner. In Fig. 2d, the dots are the polarization calculated from $\Delta\rho$ under various $r$, while the curve is a plot of Eq. (1), showing excellent agreement.

For the twisted interface, $\Delta\rho$ and $P$ as functions of $r$ can then be mapped to the dependences of these quantities on location $R$, by the local approximation. We examine the electrostatic potential at the h-BN surface produced by such a spatially modulated electrical polarization. Setting the interface to be the $z = 0$ plane, the Coulomb potential is given from the differential charge $\Delta\rho$,

$$V(\mathbf{R}, z) = \int \frac{\Delta\rho(\mathbf{R}', z')}{4\pi\varepsilon_0 \sqrt{|\mathbf{R} - \mathbf{R}'|^2 + (z - z')^2}} d\mathbf{R}' dz' \cong \text{sgn}(z) \frac{P(\mathbf{R})}{2\varepsilon_0} e^{-G|z|}, \quad (2)$$

where $\Delta\rho(\mathbf{R},z) \equiv \Delta\rho(\mathbf{r}(\mathbf{R}),z)$, $P(\mathbf{R}) = \int z'\Delta\rho(\mathbf{R},z')dz'$, and $G = \frac{4\pi}{\sqrt{3}b}$. The approximation in Eq. (2) holds well under the condition $Gd \ll 1$ (see supplementary materials). While the uniform polarization in lattice-matched stacking can only produce voltage drop between the two sides of the layer, the spatially modulated polarization at the twisted interface can introduce in-plane variation of electrostatic potential, with a strength that decays with the vertical distance $z$ from the interface. In the limit of $G|z| \ll 1$, the potential strength $V_{max} - V_{min} = \frac{1}{\varepsilon_0}P_{max}$, where $P_{max}$ is the electrical polarization at $r = a/\sqrt{3}$, i.e. AB-stacking (c.f. Fig. 2d).

Fig. 2e plot the spatial profile of electrical polarization $P(\mathbf{R})$, assuming the interface has a rigid twist from parallel, and the resultant electrostatic potential at three values of the vertical distance $z/b$. Taking $b = 9.6$ nm which corresponding to twisting angle $\delta = 1.5°$, these corresponds to the thickness of n = 2, 4 and 6 layers respectively. The choice of twisting angle and the thickness of the top h-BN layer together allows the control of both the periodicity and the strength of the superlattice potential. At the thin limit $G|z| \ll 1$, the potential strength measured by $V_{max} - V_{min}$ reaches 203 meV.

When a general 2D material is placed on the h-BN surface, the electrons/holes experience the electrostatic potential through their charge, which shall produce a universal superlattice energy landscape determined by the h-BN interface, regardless of the crystal type and orientation of the target materials. This is corroborated by our first principle calculations of the band edge energies in three distinct examples, i.e. MoSe$_2$, black phosphorus (BP), and the antiferromagnetic MnPSe$_3$ monolayers.

Fig. 3a and 3b show the DFT calculated band edges, in a monolayer MoSe$_2$ placed on an h-BN bilayer, as functions of stacking registries. In these calculations, the MoSe$_2$ has the crystal axes aligned with those of h-BN with a 4:3 ratio in their lattice constants. The periodic cell used in the calculations is shown as inset of Fig. 3b, where the interlayer translation $\mathbf{r}_t$ characterizes the registry between MoSe$_2$ and h-BN. The two h-BN layers have parallel alignment characterized by registry $\mathbf{r}_{bn}$ (inset of Fig. 3a). Fig. 3a plots the energies of conduction band minima (CBM) and valence band maxima (VBM) as functions of $\mathbf{r}_{bn}$, which shows the expected modulation by the $\mathbf{r}_{bn}$ dependent electrical polarization in h-BN. For comparison, the energies of Q and Γ valleys are also shown. Indeed, all these energy extrema are modulated with the same profile as functions of $\mathbf{r}_{bn}$. In the presence of a small twist of the h-BN interface, this dependence, together with the mapping between the local registry $\mathbf{r}_{bn}$ and location $\mathbf{R}$, produces the superlattice energy landscape for electrons in MoSe$_2$, which has the same profile for K, Q and Γ valleys.

It is generally expected that the interface between h-BN and target material such as MoSe$_2$ does not change the energy landscape of the latter, which underlies the extensive use of h-BN as substrate and capping layer. This is confirmed in our calculations, as shown in Fig. 3b, where the band edge energies show no visible dependences on the stacking registry $\mathbf{r}_t$ between MoSe$_2$ and h-BN. It is worth noting that there is a charge transfer between MoSe$_2$ and h-BN (c.f. supplementary materials), but its dependence on the registry $\mathbf{r}_t$ is smoothed out because of the large mismatch in their lattices. As the residue effect, we observe a small variation upper bounded by a few meV in MoSe$_2$'s band edge energies as functions of $\mathbf{r}_t$. We have also performed calculations for a different angular alignment between MoSe$_2$ and h-BN that forms $\sqrt{7} \times \sqrt{7}$ periodic cell, which points to the same conclusion (supplementary figure S3).

Fig. 3c and 3d show the DFT calculated CBM and VBM energies in monolayer MnPSe$_3$ with the Neel type antiferromagnetic order, placed on the bilayer h-BN. Fig. 3e and 3f are for BP monolayer on the bilayer h-BN substrate. We find the band energies in these materials are also modulated with the same profile as functions of the stacking registry $\boldsymbol{r}_{\text{bn}}$ between the two-parallel h-BN layers, while no dependence on the registry $\boldsymbol{r}_{\text{t}}$ (between h-BN and target material) is visible in this energy scale. The periodic cells used and more details of the calculations are given in the supplementary materials.

These results are in full support of the expectation that the laterally modulated electrical polarization at the twisted interface inside h-BN substrate or capping layer can generate universal superlattice energy landscape for carriers in a general 2D material. For the various band edges in the three different monolayers, the magnitude of the energy modulation falls in the range between 193 and 233 meV (c.f. Fig. 3a, 3c, 3e). The finite differences in the energy modulation strength can be attributed to the different extension of the band edge wavefunction in the z direction, as well as the dielectric screening effect in the heterostructures.

**Multiple twisted parallel interfaces in h-BN.** We consider first an AB stacking trilayer which has two parallel interfaces both with stacking registry $r = a/\sqrt{3}$. The bottom panel of Fig. 4a presents the charge redistribution due to the interlayer coupling at both interfaces, $\Delta\rho \equiv \rho_{3L} - \rho_{\text{t}} - \rho_{\text{m}} - \rho_{\text{b}}$, where $\rho_{3L}$ is trilayer charge density from DFT, and $\rho_{\text{t/m/b}}$ is that of the pristine top/middle/bottom monolayer. Through the comparison with the charge redistribution due to a single interface (c.f. top and middle panels of Fig. 4a), it is remarkable to note that $\Delta\rho$ in the trilayer is simply the sum of that due to each individual interface, showing that charge redistribution at adjacent interfaces can be treated as independent of each other. Fig. 4b plots the electrical polarization $P$ in trilayers where the two parallel interfaces have identical registry $r$. Under the various $r$, the trilayer $P$ is all twice of that in the corresponding bilayer.

The above findings point to the possibility to design superlattices from the interference of electrostatic potentials from the electrical polarizations at multiple twisted parallel interfaces. For example, creating two independently configured twisted interfaces in h-BN can allow the generation of bichromatic superlattice potentials.[36] Here we consider the scenario of staring from an $N$-layer film of AB stacking,[28] and apply torsion such that adjacent layers are all twisted from each other by a common angle $\delta$. This creates $N - 1$ nearly parallel interfaces with the same periodicity, but with a small rotation by $\delta$ from each other. With the charge redistribution at different interfaces being independent of each other, the electrostatic potentials contributed by adjacent interfaces are related by: $V_{l+1}(\boldsymbol{R}, z) = V_l(\mathbb{R}\boldsymbol{R}, z + d)$, where $\mathbb{R}$ denotes the in-plane rotation by $\delta$, with $V_l(\boldsymbol{R}, z)$ described by Eq. (2). The overall electrostatic potential is then,

$$V_{\text{sum}}(\boldsymbol{R}) = \sum_l V_0(\mathbb{R}^l \boldsymbol{R}, z + ld), \qquad (3)$$

In the limit $N\delta \ll 1$, $V_{\text{sum}}(\boldsymbol{R}) \approx \sum_l V_0(\boldsymbol{R}, z + ld)$, one obtains a superlattice potential with the strength scaled up with the thickness $N$. In Fig. 4d, we show a plot of Eq. (3) at the surface of the h-BN film for $N = 10$ and $\delta = 0.5°$. The superlattice potential strength reaches ~ 1.3 eV. Fig. 4c characterizes the potential strength with the increase of $N$. For larger $\delta$, the potential strength saturates at smaller $N$. It is interesting to note that $V_{\text{sum}}(\boldsymbol{R})$ always saturates with a common spatial profile with a chiral structure, as shown in Fig. 4e for the example of $N = 20$ and $\delta = 1°$.

**Electric field on h-BN surface and superlattice for dipolar objects.** As shown in Eq. (2), the potential $V(\mathbf{R}, z)$ decays exponentially in the out-of-plane direction in a length scale $G^{-1} = \frac{\sqrt{3}}{4\pi}b$. This is associated with an out-of-plane electric field on the h-BN surface with magnitude inversely proportional to superlattice period $b$. Fig. 5a shows the electric lines of forces which have a common profile when plotted in the dimensionless coordinates $(\mathbf{R}/b, z/b)$. For the example of a single parallel interface with $\delta = 2°$, which corresponding to $b = 7.2$ nm, the spatial profile of $E_z$ and total field strength $|\mathbf{E}|$ are shown in Fig. 5b, along the long diagonal of a supercell, at a vertical distance of $z = 2d$ from the interface. The field strength can reach $\pm 6$ mV/Å. This sizable electric field further enables the possibility to create electrostatic superlattice potential for dipolar objects such as interlayer excitons in bilayer TMDs, and dipolar molecules. Taking interlayer exciton as an example, it couples to $E_z$ through its out-of-plane electric dipole, and the superlattice potential strength in the field of Fig. 5b can reach ~ 80 meV, assuming the interlayer distance of 7 Å in TMDs. Dipolar molecules adsorbed could have their orientations adjusted to the total field $\mathbf{E}$, then the potential landscape is determined by $|\mathbf{E}|$. The electric field strength can also be scaled up by having multiple twisted parallel interfaces in h-BN, for creating a quasi-periodic disordered landscape for the dipolar object, as shown in Fig. 5c.

**Discussion**

We have shown the possibility to introduce a superlattice energy landscape on a general 2D material by exploiting twisted parallel interfaces inside the widely used h-BN substrate. Unlike the existing moiré superlattices, this non-invasive approach can be applied to the h-BN encapsulated high quality monolayers, bilayers, etc., not limited by the lattice type and periodicity of the target materials. It is also worth noting that the orientation of the superlattice potential is determined by the h-BN interface only, and can be placed at an arbitrary angle to the crystalline axis of the target material, which is also a new control parameter to tailor the minibands through various zone folding schemes. The approach can be further combined with the existing moiré superlattices in twisted graphene, twisted TMDs and TMDs heterobilayers. In addition to the substrate, the twisted parallel interfaces can also be created in the h-BN capping layer where similar effect is expected, or even in the h-BN spacing layers exploited as tunnelling barriers. The latter could point to interesting superlattice effects on the tunnelling processes.

In reality, structural relaxation can be relevant at the interfaces when the twisting angle is small.[30-32] The structural relaxation can increase the area of certain energetically favourable stacking while shrink the unfavourable ones, at the cost of inhomogeneous strain. This changes the mapping function $\mathbf{r}(\mathbf{R})$ in a supercell, and can quantitatively change the spatial profile of $P(\mathbf{R})$ from the rigid twist one shown in Fig. 2e. Correspondingly, the spatial profile of the electrostatic potential can be modulated by such structural relaxation. As long as the local approximation is still applicable, the electrical polarization profile can still be quantified using $P(\mathbf{r}(\mathbf{R}))$ from the relaxed texture $\mathbf{r}(\mathbf{R})$ of the interface. On the other hand, for the multilayers subject to torsion (Fig. 4 inset), the structural relaxation can be complicated with the large number of twisted interfaces. Compared to those shown in Fig. 4c-e for the rigid interfaces, the potential profile can get much more complex and the scaling of the potential strength with $N$ can be significantly changed by such relaxation in the small $\delta$ limit.

**Acknowledgments:** We thank helpful discussions with Xiaodong Xu, Hongyi Yu, Mingxing Chen and Qingjun Tong. The work is supported by the University of Hong Kong (Seed Funding for Strategic Interdisciplinary Research).

# Figures

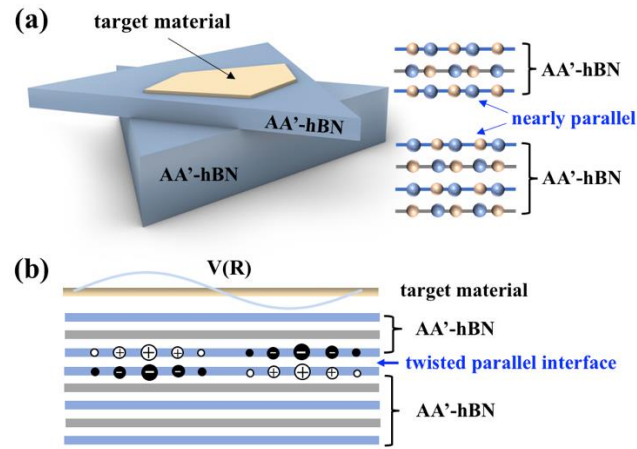

**Figure 1. Creating superlattice potential in a general 2D material from the twisted h-BN substrate**. (a) A moiré interface of nearly parallel alignment is created inside the h-BN substrate by stacking a *n*-layer AA'-hBN on another AA'-hBN film. (b) Spontaneous charge redistribution at this interface leads to electrical polarization patterned with periodicity controlled by twisting angle δ, which produces an electrostatic superlattice potential $V(\boldsymbol{R})$ in the target material with strength controlled by *n* and δ.

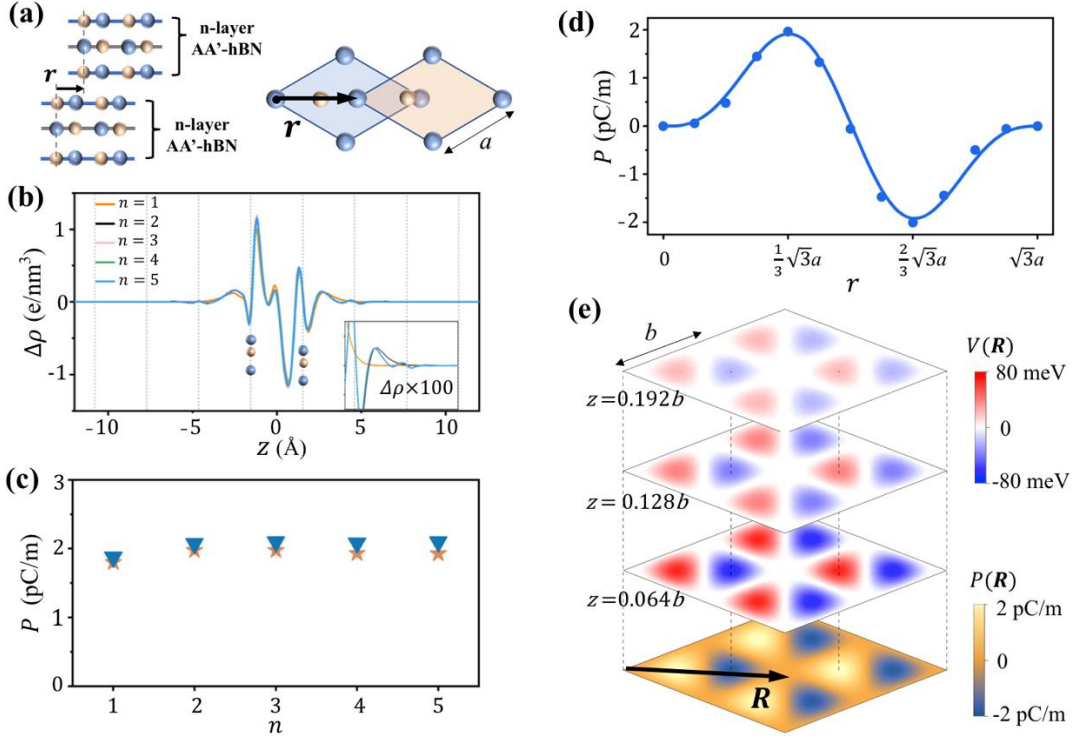

**Figure 2. Characterization of the electrostatic superlattice potential.** (a) Schematic of a $n+n$ multilayer, consisting of two n-layer AA' hBN stacked with a parallel interface. The interface is characterized by the in-plane translation $r$ between the two parallel BN layers. (b) Charge redistribution due to interlayer coupling at the parallel interface with registry $r = a/\sqrt{3}$. $\Delta\rho \equiv \rho_{n+n} - \rho_{n,t} - \rho_{n,b}$, where $\rho_{n+n}$ is the plane-averaged charge density of the $n+n$ multilayer calculated using DFT, and $\rho_{n,t/b}$ is that of the pristine top/bottom n-layer. Dashed vertical lines indicate atomic planes. For various thickness n considered, $\Delta\rho$ is predominantly distributed with nearly the same profile on the two layers forming the interface. (c) Electrical polarization in the $n+n$ multilayers, evaluated using $\int z\Delta\rho(z)dz$ (stars), and using $\varepsilon_0\Delta\varphi$ (triangles), $\Delta\varphi$ being the vacuum level difference on the two sides of multilayer directly extracted from DFT. (d) $P$ as a function of $r$ in the $2+2$ configuration. The dots denote the values from the DFT calculated $\Delta\rho$, and the curve is the fitting using Eq. (1). (e) Laterally modulated polarization $P(\mathbf{R})$ when the interface has a rigid twist from parallel by small angle $\delta$, and the resultant electrostatic superlattice potential $V(\mathbf{R}, z)$ on constant $z$ plane (from Eq. (2), c.f. text). $V(\mathbf{R})$ is shown at three vertical distance from the interface which, for the example of $\delta = 1.5°$(period $b = 9.6$ nm), corresponding to thickness of n = 2, 4, and 6 layers respectively.

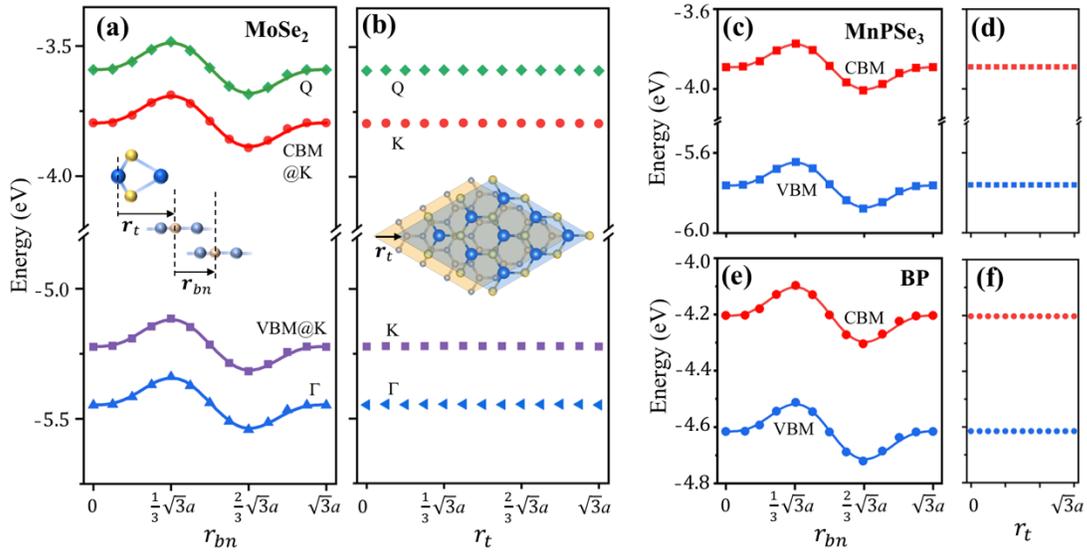

**Figure 3. Universal modulation of band-edge energies in target materials by the electrical polarization in h-BN substrate.** (a,b) Monolayer MoSe$_2$ on a bilayer h-BN substrate. From DFT calculated band structures, the energies of conduction band minima (CBM) and valence band maxima (VBM) in MoSe2 are examined as functions of $r_{bn}$ (registry between the two parallel BN layers, c.f. inset of (a)), and $r_t$ (stacking registry between MoSe$_2$ and h-BN, c.f. inset of (b)). The expected modulation by the $r_{bn}$ dependent electrical polarization in h-BN is seen in (a). The magnitude of the modulation, i.e. $E_{max} - E_{min}$, is 193 meV for K and Q valleys, and 202 meV for Γ point. The dependence on $r_t$ is not visible in (b). The energies of Q and Γ valleys show the same behaviors as the CBM and VBM @ K valleys. (c, d) and (e, f) are similar plots for MnPSe$_3$ and BP monolayer respectively on the bilayer h-BN substrate. In (c), $E_{max} - E_{min}$ is 233 meV for CBM and 223 meV for VBM. In (e), $E_{max} - E_{min}$ is 195 meV for CBM and 207 meV for VBM. See supplementary materials for more details.

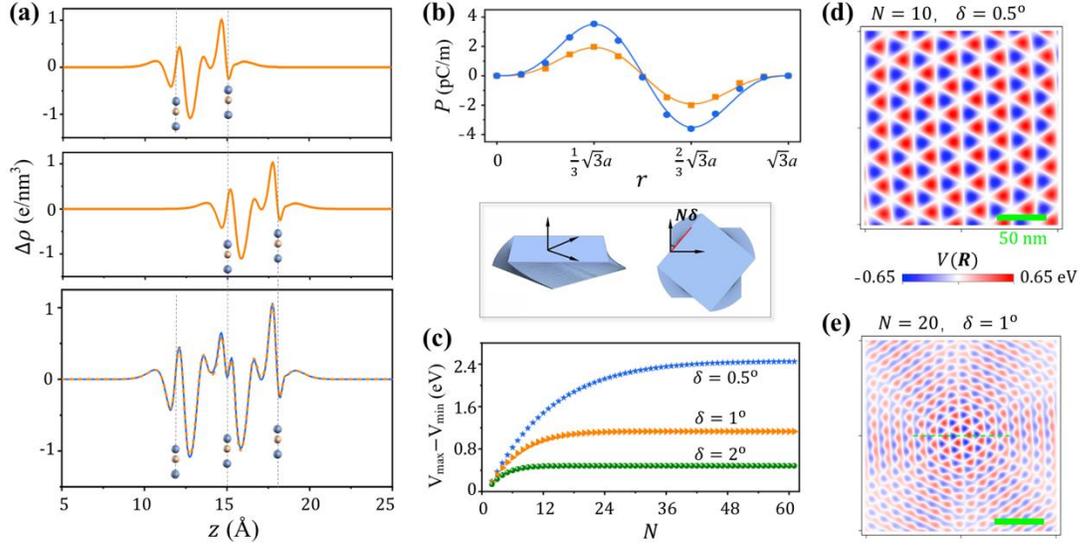

**Figure 4. Scaling up the superlattice potential using AB h-BN film subject to torsion.** (a) Top panel: charge variation $\Delta\rho$ due to interlayer coupling in AB bilayer. Middle: same as top, with shifted atomic planes. Bottom: charge redistribution due to interlayer coupling in AB trilayers. Blue curve plots $\Delta\rho \equiv \rho_{3L} - \rho_t - \rho_m - \rho_b$, where $\rho_{3L}$ is trilayer charge density from DFT, and $\rho_{t/m/b}$ is that of the pristine top/middle/bottom monolayer. Orange curve is the sum of two curves in top and bottom panels. (b) Blue dots show electrical polarization in lattice-matched trilayers with two parallel interfaces of the identical registry $r$, and orange squares show that in bilayer hBN. (c) Strength of superlattice potential in $N$-layer film under torsion where adjacent layers are all twisted with a common angle $\delta$ (c.f. inset). (d, e) Spatial profiles of superlattice potentials on the surface of 10-layer film with $\delta = 0.5°$, and 20-layer film with $\delta = 1°$. A chiral structure is seen in the latter. Green scale bar is 50 nm.

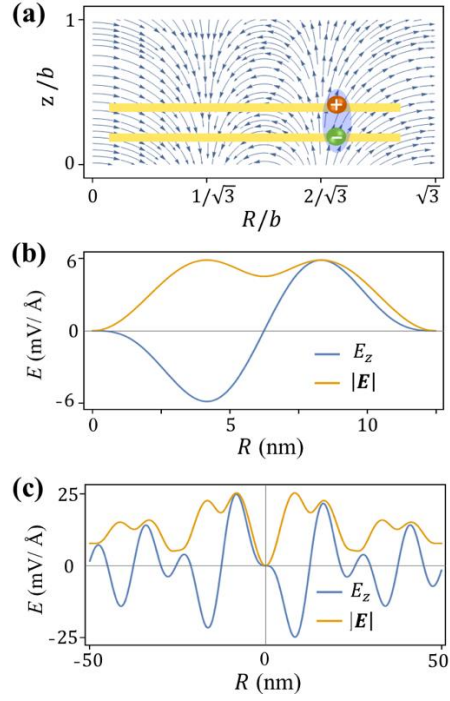

**Figure 5. Electric field on the h-BN surface.** (a) Electric lines of forces from the laterally modulated polarization $P(\mathbf{R})$ of a single twisted parallel interface with periodicity $b$ at $z = 0$. Interlayer exciton in a bilayer on the h-BN surface can thus experience a superlattice potential from the lateral modulation of out-of-plane field $E_z$. (b) Spatial profile of $E_z$ and total field strength $|\mathbf{E}|$ along the long diagonal of a supercell, at a vertical distance of $Z = 2d$ from a single interface with $\delta = 2°$. (c) $E_z$ and $|\mathbf{E}|$ on the surface of a 20-layer film under torsion with $\delta = 1°$, along the dashed line in Fig. 4(e).